\documentstyle[12pt]{article}
\newcommand{\be}{\begin{equation}}
\newcommand{\ee}{\end{equation}}

\renewcommand{\textheight}{21 cm}
\renewcommand{\textwidth}{15 cm}
\title{
   Solution of the master equation for Bak-Sneppen model
    of biological evolution in finite ecosystem
}

\author{ Yu. M.Pis`mak \\
Department of Theoretical Physics\\
Saint-Petersburg State University\\
Ul'yanovskaya 1 Stary Petergof 198904 Saint-Petersburg Russia\\
E-Mail pismak@phim.niif.spb.su
}
\date{}
\begin{document}
\maketitle
\vskip25mm
\centerline
{\large ABSTRACT}
\vskip5mm
{\small
    The master equations describing processes of biological evolution
    in the framework of the random neighbor Bak-Sneppen model are
    studied. For the eqosystem of $N$ species they are solved
    exactly and asymptotical behavior of this solution for large $N$ is
    analyzed.
}

\newpage

    \section {Introduction}
    The  model of biological evolution
    proposed by Bak and Sneppen
    [1,2] describes mutation and natural selection of interacting
    species. It is the dynamical system that is defined as follows.
    The state of the ecosystem of $N$ species is characterized
    by a set $\{ x_1,...,x_N\}$ of $N$ number, $0 \geq x_i \geq 1$.
    In so doing, $x_i$ represents the barrier toward father evolution of
    the species. Initially, each $x_i$ is set to a randomly chosen
    value. At each time step the barrier $x_i$ with minimal value and
    $K-1$ other barriers are replaced by $K$ new random numbers.
    In the random neighbor model (RNM), which
    will be considered in this paper the $K-1$ replaced
    non-minimal barriers are chosen at random.

      The RNM is the simplest model describing the avalanche-like
    processes, which are supposed by a conception of "punctuated
    equilibrium" in biological evolution. These processes are
    the most characterizing feature for self-organized criticality
    recently intensively investigated both numerically
    and analytically [3-6] The RNM is more convenient for
    analytical studies. The master equations obtained in [3]  for
    RNM are very useful for this aim. In [7] the explicit
    solution of master equations was found for infinite
    ecosystem. In this paper we solve the master equations for
    finite number $N$ of species in ecosystem. We restrict ourselves
    with the simplest case $K=2$.

   \section{Master equations for RNM}
   The master equations for the RNM are obtained in [3]. They are
   of the form:
      \begin{eqnarray}
      P_n(t+1) = A_nP_n(t) + B_{n+1} P_{n+1}(t) +
      C_{n-1}P_{n-1}(t) + \nonumber \\ D_{n+2}P_{n+2}(t) +
      (B_1\delta_{n,0}+A_1\delta_{n,1}+C_1\delta_{n,2})P_0(t)
      \end{eqnarray}
   Here, $P_n(t)$ is the probability that $n$at the time $t$ is the
   number of barriers having values less than a fixed value $\lambda$
   at the time $t$; $0 \leq n \leq N$,  $0 \leq \lambda \leq N$,
   , $0 \leq t$;
   $P_n(0)  $ are proposed to be given.
   For   $ 0<n \leq N $
   \begin{eqnarray}
   A_n = 2\lambda(1-\lambda) + \frac{n-1}{N-1}\lambda(3\lambda-2),
   \nonumber\\
   B_n = (1-\lambda)^2 +\frac{n-1}{N-1}(1-\lambda)(3\lambda-1),
   \nonumber\\
   C_n = \lambda^2 - \frac{n-1}{N-1}\lambda^2,\;\;
   D_n =(1-\lambda)^2\frac{n-1}{N-1},
   \end{eqnarray}
   and $A_n = B_n = C_n = D_n
   = 0 $ for $n=0,n>N$ .

    In virtue of the definition of $P_n(t)$,
    \be
     P_n(t) \geq 0,
    \ee
    \be
    \sum_{n=0}^{N}P_n (t) =1.
    \ee
    Making summation in (1) over $n$ and taking into account (2) it is
    easy to establish that

    \be
    \sum_{n=0}^{N}P_n (t+1) = \sum_{n=0}^{N}P_n (t).
    \ee

    Therefore, if $P_n(0)$  are chosen in such a way that (4) is
    fulfilled for $t=0$, then in virtue of (5) it is the case for
    the solution of (1) for $t>0$ too.
     For analysis of (1) it is convenient to introduce the generating
     function $q(z,u)$:
     \be
        q(z,u) \equiv
    \sum_{t=0}^{\infty}
    \sum_{n=0}^{N}P_n (t)z^n u^t.
    \ee
    In virtue of (3),(4) $q(z,t)$ is polynomial in $z$, analytical
    in $u$ for $|u|<1$ and
    \be
       q(1,u)= \frac{1}{1-u}.
    \ee
    The master equations (1) can be rewritten for the generating function
    $q(z,u)$ as follows:
    \begin{eqnarray}
    \frac{1}{u}(q(z,u)-q(z,0)) = (1-\lambda + \lambda z)^2
    \left(\frac{1}{z}\left(1- \frac{1-z}{N-1}\left(\frac{1}{z}-
    \frac{\partial}{\partial z}\right)\right)\right.\times\nonumber \\
    \times (q(z,u) -q(0,u)) + q(0,u) \Bigg).
    \end{eqnarray}
    The function $q(z,0)=\sum_{n=0}^{N}P_n (0)z^n $ in (8)
    is assumed to be given.

    \section {Asymptotic expansion of $q(z,u)$ for big $N$}
      If the function $q(z,0)$ has an asymptotic expansion in the
      region of big $N$ of the form:
      $$
         q(z,0) = \sum_{k=0}^{\infty}\frac{q_k(z,0)}{(N-1)^k}
      $$
      then the equation (8) enables one to obtain  the
      similar asymptotic expansion for $q(z,u)$:
      $$
         q(z,u) = \sum_{k=0}^{\infty}\frac{q_k(z,u)}{(N-1)^k}.
      $$
      The main approximation of $q_0(z,u)$, the function
      $
      q(z,u)
      $
      can be found from the equation
      \be
      (z-u(1-\lambda+\lambda z)^2)q_0(z,u)= zq_0(z,0) +
      u(1-\lambda+\lambda z)^2(z-1)q_0(0,u)
      \ee
      following from (8). Since $q_0(z,u)$ is analytical for
      $|z|<1, |u|<1$,
      \be
      0 = \alpha q_0(\alpha,0) +
      u(1-\lambda+\lambda \alpha)^2(\alpha-1)q_0(0,u)
      \ee
      where
      \be
      \alpha = \alpha(u) = \frac{1-2\lambda(1-\lambda)u
      - (1-4\lambda(1-\lambda u))^{\frac{1}{2}}}{2\lambda^2 u}
      \ee
      is the solution of equation $\alpha - u(1-\lambda +\lambda
      \alpha)^2 =0$.
      Obviously, for sufficient little $|u|$, $|\alpha|<1$. Thus,
      from equation (10) the function $q_0(0,u)$ can be found:
      \be
      q_0(0,u) = \frac{q_0(\alpha,0)}{1-\alpha}.
      \ee
      Substituting (12) in the right hand side of (9), one can find
      the solution in the following form:  \be q_0(z,u)
      =\frac{zq_0(z,0)(1-\alpha)+(z-1)u(1-\lambda+\lambda z)^2
      q_0(\alpha,0)}{(z - u(1-\lambda+\lambda z)^2)(1-\alpha)},
      \ee
      where $\alpha(u)$ is defined by (10).

      For $k>0$ the functions $q_k (z,u)$ are defined by recurrent
      relations
      \begin{eqnarray}
      q_k(z,u) = \frac{u(1-\lambda + \lambda z)^2(z-1)q_k(\alpha ,0 )
      + (1-\alpha)zq_k(z,0)}{(z - u(1-\lambda +\lambda z)^2)(1-\alpha)}+
      \nonumber\\
      + \frac{u(1 -\lambda + \lambda z)^2
      (r_{k-1}(z,u)+ r_{k-1}(\alpha,u))}{z - u(1- \lambda + \lambda z)^2}.
      \end{eqnarray}
      Here,
      \be
         r_k(z,u) \equiv z\frac{\partial}{\partial z}
         \frac{q_k(z,u)- q_k(0,u)}{z}.
      \ee
      In virtue of (13), (14) the first correction to the lowest
      approximation (12) of $q(z,u)$ has the form
      \begin{eqnarray}
      q_1(z,u) = \frac{u(1-\lambda + \lambda z)^2(z-1)q_1(\alpha ,0 )
      + (1-\alpha)zq_1(z,0)}{(z - u(1-\lambda +\lambda z)^2)(1-\alpha)}
      +\nonumber \\
      + \frac{u(1 -\lambda + \lambda z)^2
      (r_{0}(z,u)+ r_{0}(\alpha,u))}{z - u(1- \lambda + \lambda z)^2}.
      \end{eqnarray}
       where
       \be
         r_0(z,u) = z\frac{\partial}{\partial z}
         \frac{(1-\alpha)q_0(z,u)+ (u(1- \lambda + \lambda z)^2 - 1)
          q_0(\alpha,0)}
      {(z-u(1- \lambda + \lambda z)^2)(1-\alpha)}.
      \ee

      \section{Exact form of $q(z,u)$}

      Let us introduce the quantity
       \be
         Q(z,u) \equiv
         \frac{q(z,u) - q(0,u)}{z}.
      \ee
      It follows from (8) that this function fulfills the relation
      of the form:
      \begin{eqnarray}
      (z - u(1-\lambda + \lambda z)^2 - \frac{u(1-\lambda + \lambda z)^2
      (1-z)}{N-1}\frac{\partial}{\partial z}) Q(z,u)= \nonumber\\
     = q(z,0) + (u(1-\lambda + \lambda z)^2 - 1)q(0,u).
      \end{eqnarray}
      This inhomogeneous differential equation for $Q(z,u)$ has a
      special solution
      \be
         Q(z,u) = (N-1)e^{R(z,u)}\int\limits_z^1 e^{-R(x,u)} g(x,u)dx
         \equiv Q_{sp}(z,u).
      \ee
      Here,
      \be
      R(z,u) = \frac{N-1}{u}(\ln(1-\lambda + \lambda z)  -(1-u)
      \ln(1-z) + \frac{1-\lambda}{\lambda (1-\lambda + \lambda z) }),
      \ee
      \be
      g(x,u) = \frac {q(x,0) + (u(1-\lambda + \lambda z)^2-1)
      q(0,u)}{u(1-\lambda + \lambda z)^2(1-x)}
      \ee
      and the derivative of $R(z,u)$ with respect to $z$ has the form
      \be
       \frac{\partial R(z,u)}{\partial z} =
       \frac{(N-1)(1-(1-\lambda + \lambda z)^2)}
       {u(1-\lambda + \lambda z)^2(1-z)}
       \ee
       General solution of the corresponding to (19) homogeneous
       equation
       \be
       (z - u(1-\lambda + \lambda z)^2 - \frac{u(1-\lambda + \lambda
       z)^2 (1-z)}{N-1}\frac{\partial}{\partial z}) S(z,u)= 0
       \ee
       is of the form
       \be
       S(z,u) = F(u)e^{R(z,u)}.
       \ee
       Here, $F(u)$ is an arbitrary function of $u$.
       Hence, it follows from (19),(20),(25)
       that the function $Q(z,t)$ can be represented
       as follows:
       \be
       Q(z,u)= F(u)e^{R(z,u)} + Q_{sp}(z,u).
       \ee
       In virtue of initial condition (7) for $q(z,u)$
       \be
       Q(1,u) =
         \frac{1}{1-u} - q(0,u)
       \ee
       For $0<u<1, z\rightarrow 1 $,
       $S(z,u)$ diverges and $S_{sp}(z,u)$ has the finite limit:
       \be
        \lim\limits_{z\rightarrow 1} Q_{sp}(z,u) = \frac{1}{1-u}
        - q(0,u)
       \ee
        Hence, $F(u)=0$ in (25) and this representation for $Q(z,u)$
        can be rewritten in the form:

        \begin{eqnarray}
         Q(z,u) = (N-1)e^{R(z,u)}\int\limits_{\frac{\lambda -1}
       {\lambda}}^1
        e^{-R(x,u)} g(x,u)dx + \nonumber\\
        + (N-1)e^{R(z,u)}\int\limits_z^{\frac{\lambda -1}{\lambda} }
        e^{-R(x,u)} g(x,u)dx
       \end{eqnarray}
        It follows from (18) that
        $$ Q(\frac{\lambda - 1}{\lambda}, u) = \frac{\lambda(
        q(\frac{\lambda - 1}{\lambda}, u)- q(0,u))}{\lambda - 1}
        $$
         For the terms in the right hand side of (29) we have for
        $0<u<1$, $0<\lambda<1$:
         \be
        \lim\limits_{
        z \rightarrow \frac{\lambda -1}{\lambda} +0}
        e^{R(z,u)} = + \infty,
        \ee
        \be
        \lim\limits_{
        z \rightarrow \frac{\lambda -1}{\lambda} +0}
        (N-1)e^{R(z,u)}\int\limits_z^{\frac{\lambda -1}{\lambda}
        } e^{-R(x,u)} g(x,u)dx = \frac{\lambda(
        q(\frac{\lambda-1}{\lambda}, u)- q(0,u)}{\lambda-1}.
        \ee
       Therefore (29) can represent the function $Q(z,u)$ with
        necessary analytical properties only if
         \be
         \int\limits_{\frac{\lambda -1}{\lambda}}^1
         e^{-R(x,u)} g(x,u)dx =0
         \ee
       This equation defines the function $q(0,u)$:
         \be
         q(0,u)= \frac
         {
         \int\limits_{\frac{\lambda -1}{\lambda}}^1
         e^{-R(x,u)}\frac{q(x,0) dx}{(1-\lambda +\lambda x)^2(1-x)}
         }
         {
         \int\limits_{\frac{\lambda -1}{\lambda}}^1
         e^{-R(x,u)}\frac
         {(1-u(1-\lambda +\lambda x)^2) dx}
         {(1-\lambda +\lambda x)^2(1-x)}
         }
         \ee

         Thus, we obtain from (29) the solution of equation (8) in the
         following form:
         \begin{eqnarray}
          q(z,u) = z\frac{N-1}{u}e^{R(z,u)}\int\limits_z^{\frac{\lambda -1}
         {\lambda}}e^{-R(x,u)}\frac{q(x,0) dx}{(1-\lambda +
          \lambda x)^2(1-x)}+ \nonumber\\
          + q(0,u)(1 + z\frac{N-1}{u}e^{R(z,u)}
         \int\limits_z^{\frac{\lambda -1}
         {\lambda}}e^{-R(x,u)}\frac
         {(1-u(1-\lambda +\lambda x)^2)dx}
         {(1-\lambda +\lambda x)^2(1-x)}),
         \end{eqnarray}
         where $q(0,u)$ is defined by (33).

         \section{Conclusion}
         We constructed the solution of the master equation (8) for
         the finite number $N$ of species in the ecosystem. It can be
         proven that the main term (13)
         of its asymptotic for large $N$ coincides with the one obtained
         in [7].  Using (34) one can obtain all
         the known analytical results for RNM. One can hope that it
         helps to understand better the most important properties of
         the self-organized criticality processes.

\end{document}